\documentclass[aps,pra,twocolumn,superscriptaddress,floatfix]{revtex4}
\usepackage{amsmath}
\usepackage{amsfonts}
\usepackage{amssymb}
\usepackage{graphicx}
\usepackage{cancel}
\usepackage[switch,columnwise]{lineno}
\usepackage{titlesec}
\makeatletter
\renewcommand{\section}{\@startsection{section}{1}{0mm}
{-\baselineskip}{0.5\baselineskip}{\bf\leftline}}
\renewcommand{\subsection}{\@startsection{section}{1}{0mm}
{-\baselineskip}{0.5\baselineskip}{\bf\leftline}}
\makeatother
\begin{document}
\title{Spatial multiplexing of squeezed light by coherence diffusion}
\author{Jian Sun}%
\affiliation{Department of Physics, State Key Laboratory of Surface Physics and Key Laboratory of Micro
and Nano- Photonic Structures (Ministry of Education), Fudan University, Shanghai 200433, China}%
\author{Weizhi Qu}%
\affiliation{Department of Physics, State Key Laboratory of Surface Physics and Key Laboratory of Micro
and Nano- Photonic Structures (Ministry of Education), Fudan University, Shanghai 200433, China}%
\author{Eugeniy Mikhailov}%
\affiliation{Department of Physics, College of William and Mary, Williamsburg, Virginia 23185, USA}%
\author{Irina Novikova}%
\affiliation{Department of Physics, College of William and Mary, Williamsburg, Virginia 23185, USA}%
\author{Heng Shen}%
\email{heng.shen@physics.ox.ac.uk}
\affiliation{Clarendon Laboratory, University of Oxford, Parks Road, Oxford, OX1 3PU, UK}%
\author{Yanhong Xiao}%
\email{yxiao@fudan.edu.cn}
\affiliation{Department of Physics, State Key Laboratory of Surface Physics and Key Laboratory of Micro
and Nano- Photonic Structures (Ministry of Education), Fudan University, Shanghai 200433, China}%
\begin{abstract}
Spatially splitting nonclassical light beams is in principle prohibited due to noise contamination during beam splitting. We propose a platform based on thermal motion of atoms to realize spatial multiplexing of squeezed light. Light channels of separate spatial modes in an anti-relaxation coated vapor cell share the same long-lived atomic coherence jointly created by all channels through the coherent diffusion of atoms which in turn enhances individual channel's nonlinear process responsible for light squeezing. Consequently, it behaves as squeezed light in one optical channel transferring to other distant channels even with laser powers below the threshold for squeezed light generation. An array of squeezed light beams was created with low laser power $\sim\text{mW}$. This approach holds great promise for applications in multi-node quantum network and quantum enhanced technologies such as quantum imaging and sensing.
\end{abstract}
\maketitle
Spatially splitting a beam of classical light is a routine process to distribute the light energy in optics, and can be conveniently done by a beam splitter made of glass or crystal. However, such an operation for a quantum state of light is nontrivial since a conventional beam splitter introduces noise contamination from the vacuum, destroying the quantum property of the light.  As a consequence, when spatially separated multiple nonclassical sources are needed, such as in multi-node quantum networks~\cite{Kimble, Jia}, and studies of multi-partite entanglement for continuous variables~\cite{Su,Furusawa,PKLam,Pfister}, one has to duplicate several setups, each including a parametric nonlinear crystal and a cavity to generate one beam of squeezed light or one photon pair. How to scalably produce an array of nonclassical light beams therefore becomes a question of great interest to both fundamental studies of quantum optics and applications in quantum communications and quantum imaging with high spatial resolution~\cite{Bowen,Romalis,Giovannetti}.

In this paper, we propose a scheme and report the proof-of-principle experiment to spatially multiplex squeezed light using fast moving atoms. The essence is to employ the transport of long-lived atomic coherence within an anti-relaxation coated vapor cell, combined with coherence-enhanced nonlinear atom-light interaction. Multiple spatially separated laser beams (channels) are shined into the cell, and serve as the mutual optical pumping or state preparation for each other to create the atomic coherence, and further leads to the generation of squeezing in all optical channels. A striking phenomenon is the observation of squeezing in an optical channel with the laser power below the squeezing threshold when a distant beam is introduced. Previously, such coherence transport has been used to demonstrate anti-parity-time symmetric optics~\cite{Peng} and a slow light beamsplitter~\cite{YXiao} for classical light, with however no evolvement of the quantum properties of the optical fields.

\begin{figure}
\centering
\includegraphics[width=0.45\textwidth]{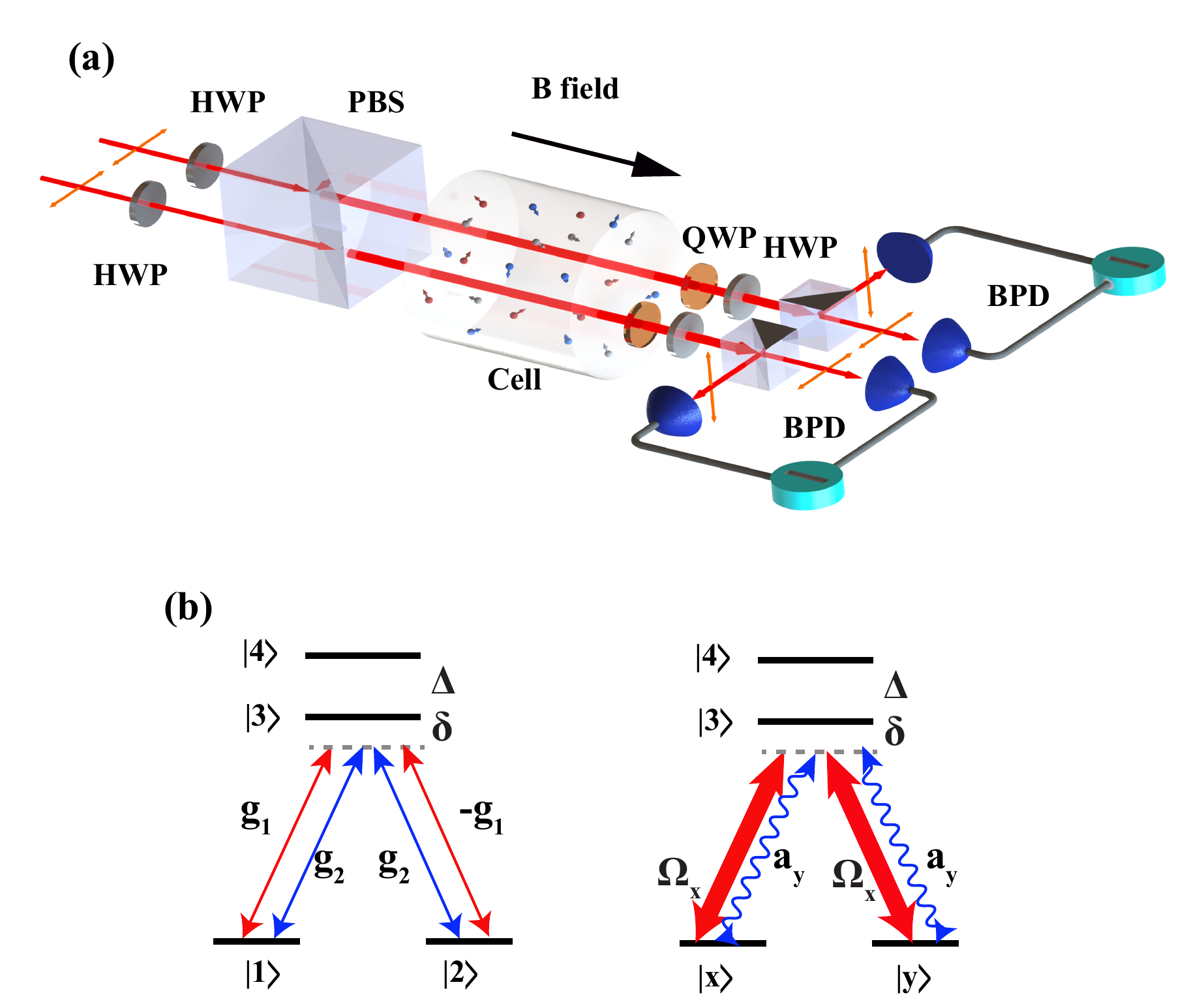}
\caption{\label{Fig:Setup} Schematics for spatial multiplexing of squeezed light.
(a) Experiment schematics. Two spatially separated optical channels (Ch1 and Ch2), formed by $x$-polarized laser beams propagating along $z$, interact with Rb atoms inside an anti-relaxation coated vapor cell, and then have their quantum fluctuations individually analyzed by balanced photodetectors (BPD) in a homodyne configuration. PBS: polarization beam splitter; HWP: Half-wave plate; QWP: Quarter-wave plate, used a phase retarder here to rotate the quantum noise ellipse. (b) The double-$\Lambda$ four-level interaction schemes. A $x$-polarized driving laser is near resonant with the $\left |5S_{1/2},F=2 \right \rangle \rightarrow \left |5P_{1/2},F'=1,2 \right \rangle$ transition. In the circular basis (left diagram), the driving laser is treated as a superposition of the left- and right-circularly polarized components, driving the dipole transitions between the Zeeman sublevels of the ground state and both excited states: $\left | 1 \right \rangle\rightarrow\left | 3,4 \right \rangle$ (with the single photon Rabi frequencies $g_1$, $g_2$ ) and $\left | 2 \right \rangle\rightarrow\left | 3,4 \right \rangle$ (with Rabi frequencies $g_2$, $-g_1$) correspondingly. The energy levels can be redrawn in the linear atomic basis (right diagram) with $\left | x \right \rangle=\frac{1}{\sqrt{2}}(\left |1 \right \rangle+\left |2 \right \rangle)$ and $\left | y \right \rangle=\frac{1}{\sqrt{2}}(\left |1 \right \rangle-\left |2 \right \rangle)$ to highlight the FWM description of the PSR. The $x$ polarization drives $\left | x \right \rangle\rightarrow\left | 3 \right \rangle$ and $\left | y \right \rangle\rightarrow\left | 4 \right \rangle$, whereas the $y$ polarization drives $\left | x \right \rangle\rightarrow\left | 4 \right \rangle$ and $\left | y \right \rangle\rightarrow\left | 3 \right \rangle$, as determined by the Clebsch-Gordan coefficients of $^{87}$Rb. The detuning of laser light from the
$\left | 1 \right \rangle\rightarrow\left | 3 \right \rangle$ transition is denoted as $\delta$, and the hyperfine splitting between the two excited states is $\Delta=\text{814.5 MHz}$.}
\end{figure}

\begin{figure}
\centering
\includegraphics[width=0.45\textwidth]{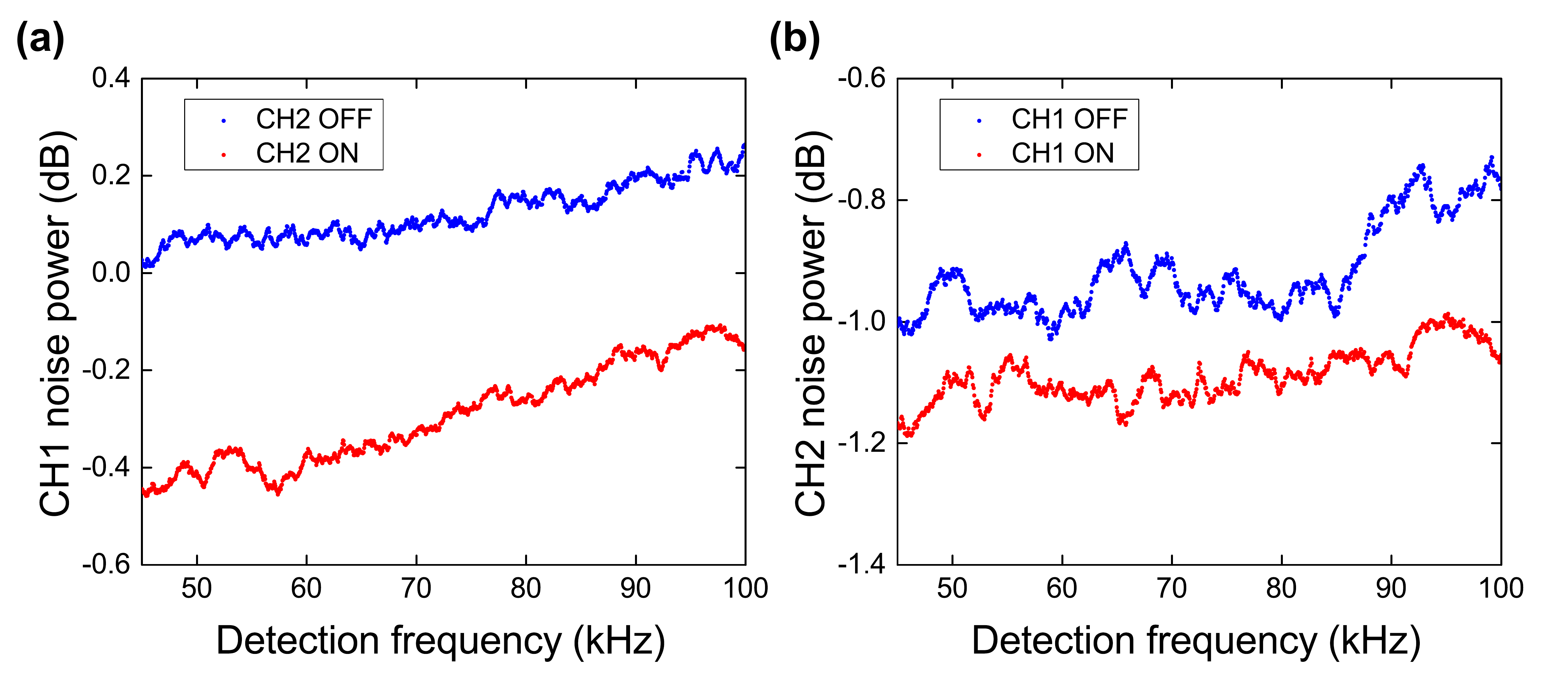}
\caption{\label{Fig 2} Proof-of-principle two-channel squeezed light result. Measured minimal-quadrature noise spectra for both channels with the other channel on or off, obtained by a spectrum analyzer. The input laser powers for Ch1 and Ch2 before the cell are $\text{4.5 mW}$ and $\text{11 mW}$, respectively. The laser frequency is $\text{348 MHz}$ blue-detuned from$^{87}\text{Rb}$ D1 line $F=2$ to $F'=2$ transition. The noise power is normalized to the shot noise level; the spectrum analyzer is set to the resolution bandwidth (RBW) $\text{10 kHz}$, video bandwidth (VBW) $=\text{10 Hz}$, each trace is averaged 200 times.}
\end{figure}

For light squeezing, we consider a double-$\Lambda$ scheme typically used for squeezed light and photon-pair generation in atom ensembles~\cite{Lett,Du} in absence of an optical cavity thanks to the atomic coherence which has been proved to enhance nonlinear interactions~\cite{RMP} and reduce laser power requirements. In our system we take advantage of the long-lived spin coherence created between Zeeman sublevels, which is known to enhance the nonlinear magneto-optical rotation and polarization self-rotation effects (PSR)~\cite{Budker}. The latter can be used to generate squeezed vacuum in an orthogonal polarization if the linearly polarized light propagates through the PSR medium~\cite{Matsko}, due to the enhanced cross-phase modulation between the left and right circularly polarized components of the $x$-polarized laser field, interacting with the optical transitions $\left | 1 \right \rangle\rightarrow\left | 3,4 \right \rangle$
and $\left | 2 \right \rangle\rightarrow\left | 3,4 \right \rangle$ as shown in Figure~\ref{Fig:Setup}(b). Alternatively, such interaction can be described as a degenerate four-wave-mixing (FWM) process where the driving laser serves as the pump twice in the FWM cycle, generating a pair of degenerate $y$-polarized photons, as shown in the energy diagram redrawn in the linear atomic basis. These quantum correlated $y$-polarized photon pairs can be treated as a quadrature-squeezed $y$-polarized vacuum, whereas the whole light field is in a polarization squeezed state~\cite{Lvovsky,Mitchell,Wasilewski,Matsko,Grangier}.

In our proposal, two (or more) illuminated interaction regions, named optical channels, and the dark region outside of the laser beams in the cell are considered (Figure~\ref{Fig:Setup}(a)), with each channel undergoing the above double-$\Lambda$ process~\cite{Rochester}. The spin dynamics of the moving atoms can be described by a set of coupled differential equations taking into account the atomic coherence exchange between different regions, as well as the Langevin noise operators. The effective coherence exchange between the optical channels is driven by the slow evolution of the ground state spin of atoms in the dark region, and the optical coherence transfer between channels is negligible as it decays within 20 ns (for $^{87}$Rb). Since FWM within either channel typically has a faster cycle than the slow spin dynamics, as implied by the much broader noise spectrum of squeezing, FWM in all channels essentially sees and is boosted by the same collective spin state, causing the simultaneous quantum noise reduction or light squeezing in all channels.

The experiment schematics is shown in Figure~\ref{Fig:Setup}. The paraffin-coated cylindrical Pyrex cell ($7.5~\text{cm}$ in length and $\text{2.5 cm}$ in diameter) contains isotopically enriched $^{87}\text{Rb}$ vapor, and is at the maximum operational temperature of $\text{55}^{\circ}$, limited by the coating. The cell was mounted inside a four-layer magnetic shielding.  A diode laser was tuned to the D1 line of $^{87}\text{Rb}$ and its output passed through a polarization-maintaining optical fiber and was separated into two or four beams. All beams passed through the same polarization beam splitter (PBS) before the cell to ensure the same polarization of all beams. All laser beams are slightly focused via a 1-meter-focal-length lens before the cell. The linearly polarized input laser also played the role of the local oscillator (LO) at the output for quantum noise measurements of squeezing~\cite{Lezama,NovikovaOL,LezamaExpt}. 

Since the average time between atom-wall collisions is about $100~\mu \text{s}$ and the transit time through the light beam with diameter of $\text{1.5 mm}$ is $\sim5~\mu \text{s}$, the long-lived ground state atomic spin (coherence lifetime about $\text{30 ms}$) can thus endure nearly a thousand wall collisions~\cite{Balabas}, which establishes within the entire cell a steady state distribution of coherence and population attributed to optical pumping by all the optical channels~\cite{Borregaard,Firstenberg}. It is then expected that turning on one channel can enhance coherence within the entire cell hence light squeezing in another channel due to effectively increased average optical pumping rate, which was verified by Figure~\ref{Fig 2}. In contrast to the nearly shot-noise-limited (SNL) output from Ch1 ($\sim \text{4.5 mW}$ input power) in absence of Ch2, squeezed light is observed below $\sim\text{100 kHz}$ when switching on Ch2 ($\sim \text{11 mW}$ input power), with $\text{0.45 dB}$ squeezing detected at $\text{40 kHz}$. Similarly, as shown, the minimal noise for Ch2 is also reduced upon turning on Ch1.

\begin{figure}
\centering
\includegraphics[width=0.5\textwidth]{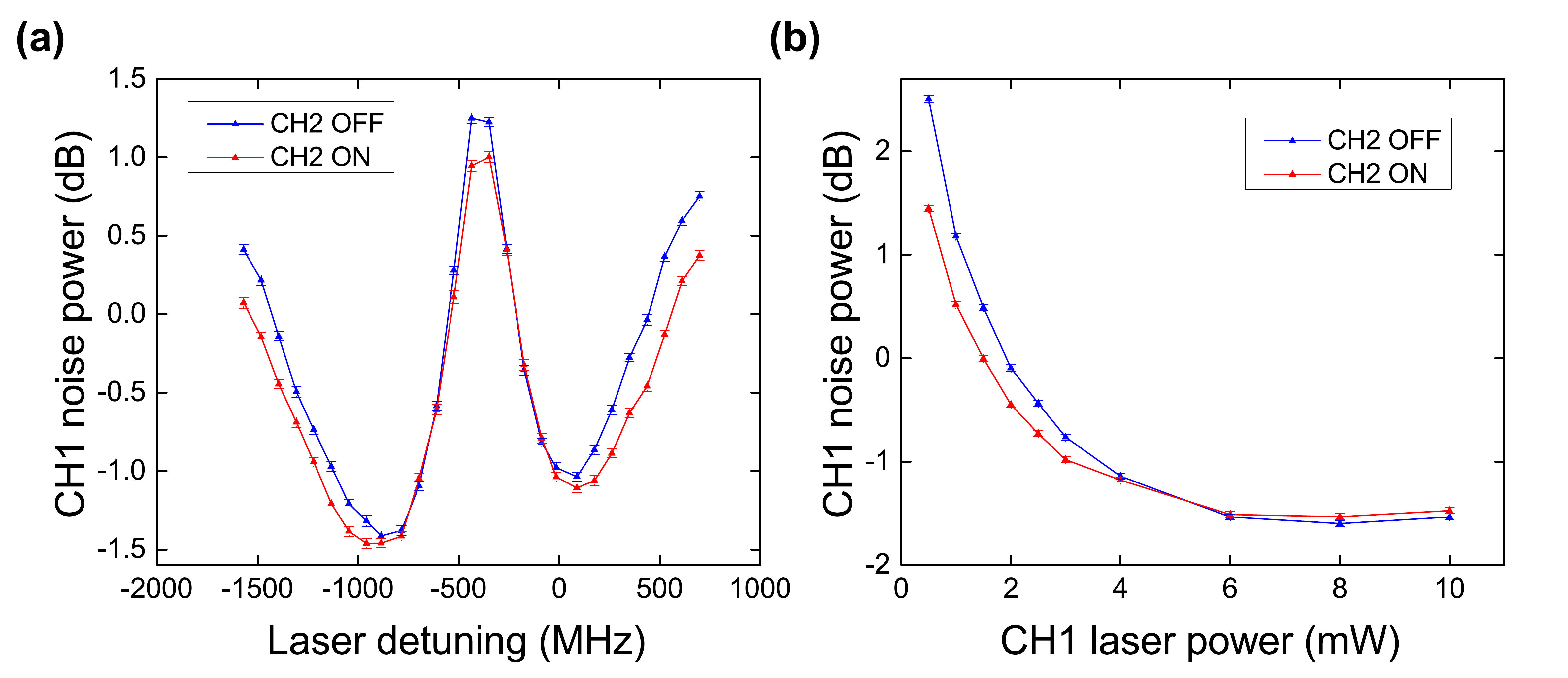}
\caption{\label{Fig 3} Dependence on the laser detuning and laser power of the noise reduction in quadrature with minimum variance. (a) Laser detuning dependence of the squeezed quadrature of Ch1 with and without Ch2. Zero detuning refers to resonance with $5S_{1/2}F=2 \rightarrow 5P_{1/2} F'=2$. The input laser power is $\text{5 mW}$ in Ch1 and $\text{6 mW}$ in Ch2. (b) Minimum quadrature noise in Ch1's output as a function of Ch1's input laser power, with and without the Ch2 laser beam of $6$~mW power, with the laser frequency $\text{56 MHz}$ red-detuned from the $5S_{1/2}F=2 \rightarrow 5P_{1/2} F'=1$ transition. Each data point is measured at the detection frequency $\text{40 kHz}$ and is averaged 100 times.}
\end{figure}

To fully characterize such noise reduction by a remote laser beam, we investigate the dependence of quantum noise on the laser detuning and laser power. For the first measurement, we fix the laser power of Ch1 at $\text{5 mW}$ and Ch2 at $\text{6 mW}$, and scan the laser frequency across the $\left |5S_{1/2},F=2 \right \rangle \rightarrow \left |5P_{1/2},F'=1,2 \right \rangle$ transitions. Figure \ref{Fig 3} shows the recorded noise floor of quadrature with minimum variance for each laser frequency. We can clearly see the quantum noise reduction in the presence of the second channel in the whole scanning range except for the small region near the transition $\left |5S_{1/2},F=2 \right \rangle$ to $\left |5P_{1/2},F=1,2\right \rangle$. We can explain this by the saturation effect that is more pronounced under parameter regimes for stronger squeezing such as higher laser power and near resonance. To verify the saturation effect related to laser power, we measure the noise reduction in Ch1 at different laser powers due to presence of Ch2 (with fixed laser power $\text{6 mW}$) under the near resonance regime, as shown in Fig.3(b), and indeed found that noise reduction decreases at higher power. Similar trend was also observed in the off-resonant regime but with higher saturation power. The saturation has two contributions. On one hand, the nonlinearity and associated ground state coherence for FWM is stronger near the resonance and at higher laser power, resulting in stronger squeezing. On the other hand, population loss into the other hyperfine ground state due to optical depumping is also more severe in the same regime, leading to reduced optical depth for squeezing. Such reduction of effective optical depth at higher laser power has been confirmed before in a similar system in the slow light studies~\cite{EIT-ABC}. The above analysis and observation has been qualitatively reproduced by our numerical simulation. The detailed model accounting for the above saturation as well as multi-level effects will be presented in a separate publication.

One important requirement for this scheme is that all channels create in-phase atomic coherence. For instance, the above noise reduction will be affected, if we vary the polarization direction of the Ch1 input driving laser (with respective to that of Ch2), which determines the phase of Ch1's prepared atomic coherence, as can be understood by the diagram in the circular basis shown in Figure~\ref{Fig:Setup}(b). In particular, an intuitive picture can be made when the laser is near resonance $5S_{1/2}F=2 \rightarrow 5P_{1/2} F'=1$ : the resonant $\Lambda$ process establishes Electromagnetically-Induced-Transparency (EIT) and prepares atoms into a dark state $g_2\left |1 \right \rangle-g_1\left |2 \right \rangle$ which relies on the relative phase of the two circular light components (i.e., light polarization direction), while the off-resonant $\Lambda$ introduces AC-stark shift only related to the two light intensities. Fig.4 shows the noise reduction in quadrature with minimum variance as a function of relative polarization angle between two laser beams. As expected, the noise reduction of detected channel is maximized with same polarization of two beams and decreases with the relative polarization angle. This result can be explained by the destruction of the state preparation process with different polarization of two beams, which pump atoms to different states, giving rise to a reduced average coherence in the cell upon motional averaging, and subsequently suppress FWM and light squeezing. The experiments performed in near-resonant and detuned regime give similar trend.

\begin{figure}
\includegraphics[width=0.7\columnwidth]{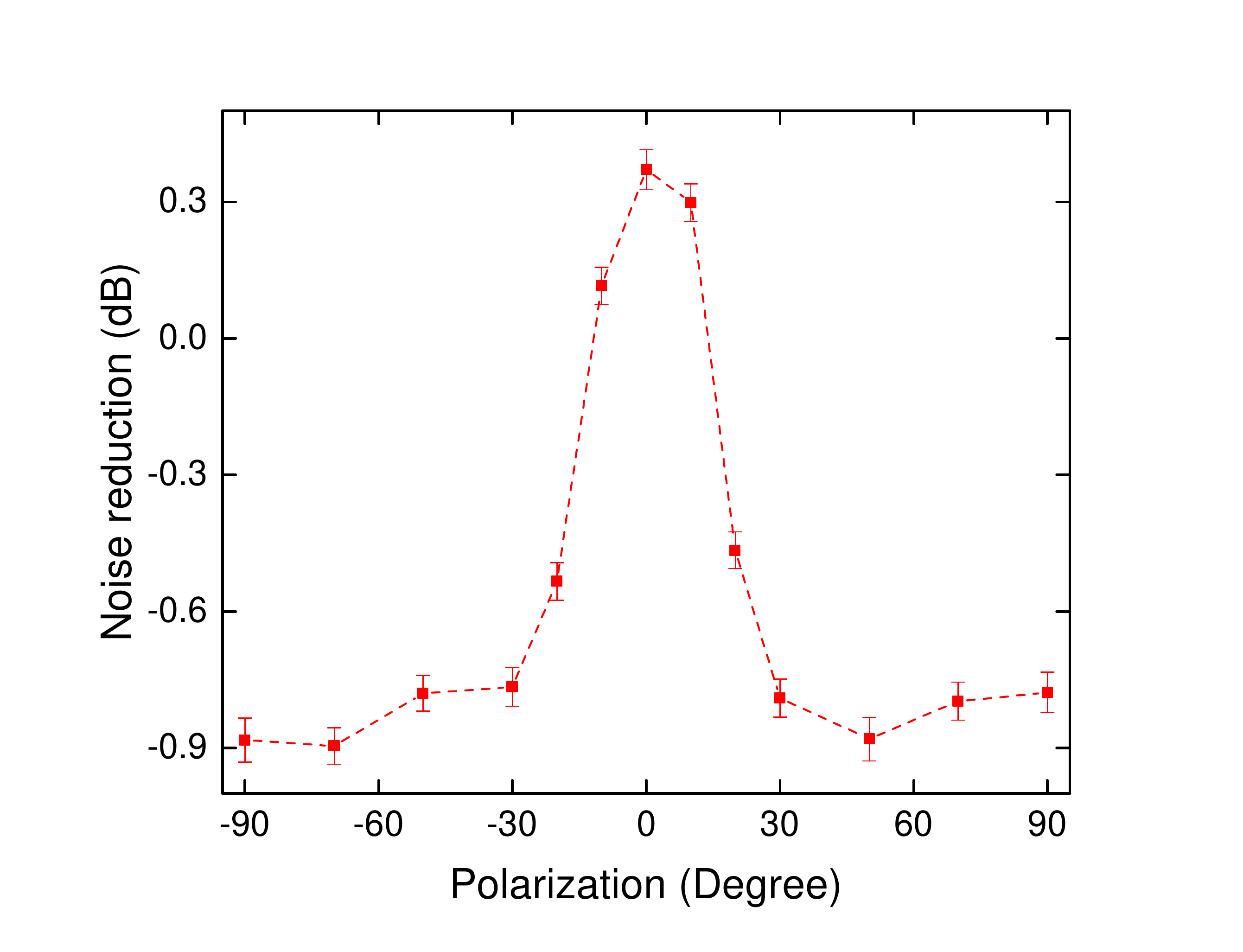}
\caption{Noise power reduction of Ch1 in quadrature with minimum variance, after turning on Ch2, as a function of the relative polarization angle between Ch1 and Ch2. The amount of noise reduction (positive value) compared to the shot noise level decreases as the polarization direction of the two channels offset from each other. Ch1's polarization is fixed along $x$ direction. The laser frequency is $\text{348 MHz}$ blue-detuned from$^{87}\text{Rb}$ D1 line $F=2$ to $F'=2$ transition. The laser power of Ch1 and Ch2 before the cell are $\text{4.5 mW}$ and $\text{6 mW}$, respectively. The spectrum analyzer settings are RBW $=\text{10 kHz}$, VBW $=\text{10 Hz}$. Each data point is averaged 200 times.}
\end{figure}

\begin{figure}
\includegraphics[width=0.8\columnwidth]{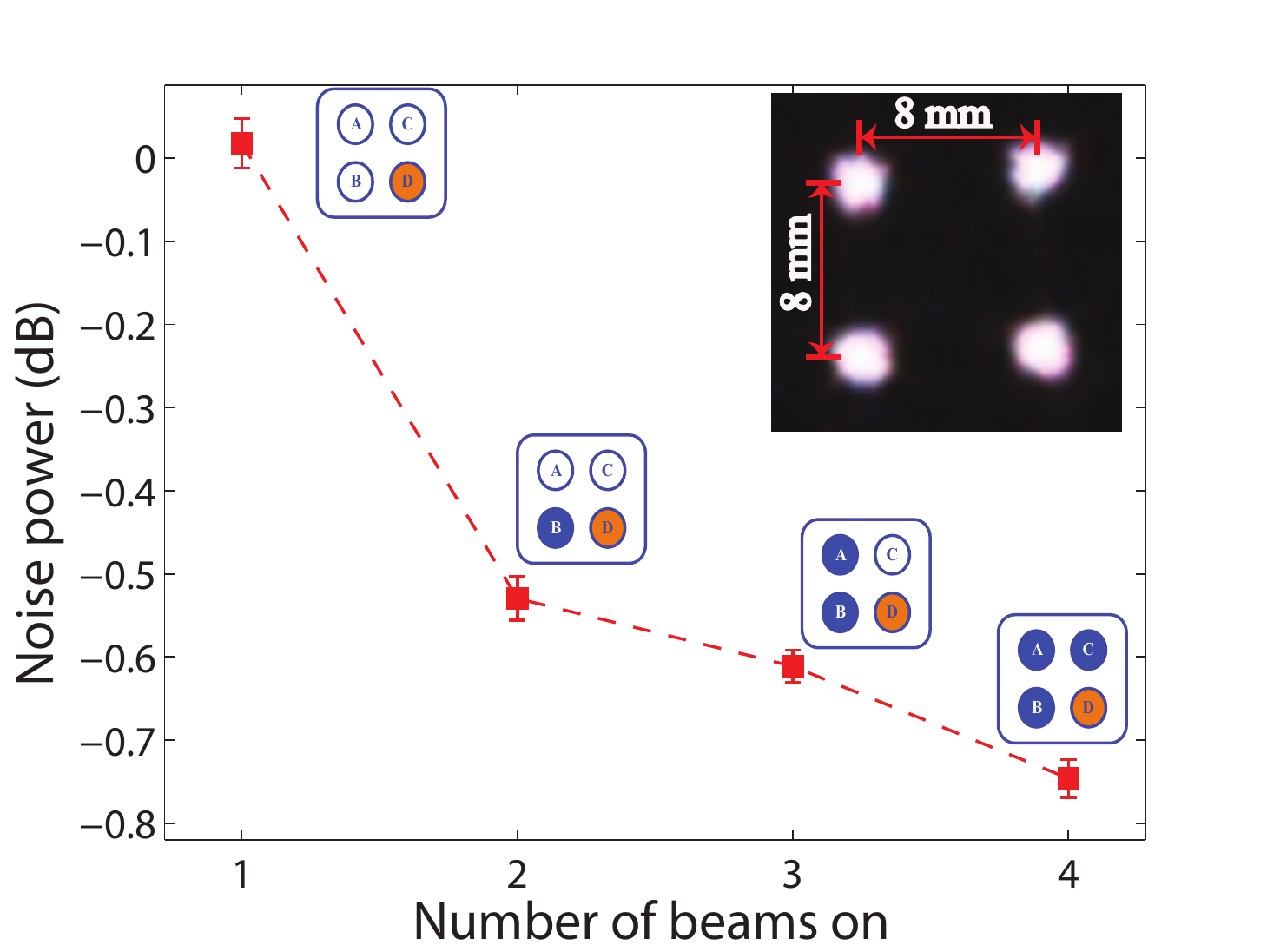}
\caption{$2\times 2$ array of squeezed light. Noise power of channel D with presence of different number of channels.
Insert shows the spatial mapping of $2\times 2$ parallel laser beams taken by CCD. The distance between two adjacent beams is $\text{8 mm}$.
The laser power of each channel before cell is $\text{4.25 mW}$. The laser frequency is $\text{348 MHz}$ blue-detuned from $^{87}\text{Rb}$ D1 line
$F=2$ to $F'=2$ transition. The noise power is relative to the shot noise level ($\text{0 dB level}$), and the spectrum analyzer settings are
RBW $=\text{10 kHz}$, VBW $=\text{10 Hz}$. Each trace is averaged 50 times.}\label{Fig6}
\end{figure}

According to the motional averaging process, it is natural to extend the one-dimensional results shown above to a two-dimensional (2D) configuration. We investigate the enhancement of squeezed light in a $2\times 2$ array (with the four channels named as A,B,C and D), and demonstrate a 2D array of squeezed light with low laser power, as shown in Figure \ref{Fig6}. We fixed all four laser beams power at $\text{4.25 mW}$ and measured the noise power in quadrature with minimum variance of Channel D. We found that, with more channels switched on, the degree of squeezing observed steadily increases. Compared to the almost SNL noise power of Channel D initially, $\sim\text{0.75 dB}$ squeezing appears when all other three channels are switched on. Furthermore, with more number of beams on in the array, the noise powers of the squeezed quadrature in all channels are lower. As shown in Figure \ref{Fig6}, to achieve the same light squeezing (0.75 dB) in all channels, each channel has laser power of $\text{4.25 mW}$ in a $2\times 2$ array, in contrast to the laser power of $\text{7 mW}$ for the single channel case at the same laser detuning. This results shows the potential to realize spatially-multiplexed array of squeezed light with lower laser power (per channel) by increasing the number of channels. We found that, the maximal achievable squeezing in each channel when all beams are on is about the same as that when only one channel is present (about 1.5 dB in this setup), when the laser power and detuning are optimized independently for both cases. The maximal squeezing is determined by the maximal coherence and optical depth which is the same for single channel and multiple channel scenario, all in the saturation regime. Finally, an even larger array of squeezed beam is immediately obtainable with a proper device for creating an array input, such as acoustic-optical deflector.

Although the degree of squeezing in the current setup is moderate, further improvements could be made mainly by increasing the effective optical depth. For instance, an additional repump beam can be applied to collect the atoms pumped into the other hyperfine ground state. The optical depth can be also enhanced by using a longer vapor cell or applying wall coatings that work at higher temperature~\cite{OTS}. In addition, engineering of the spatial mode of the light involved in this scheme can further enhance squeezing~\cite{Novikova-Downing}. Finally, the squeezing at lower detection frequency should be even higher according to the theory, and could be observed by developing proper detection schemes shifting the spectrum to higher frequency. So far, the highest single-channel polarization squeezing in atomic vapor is about 3 dB (4.7 dB after loss corection)~\cite{LezamaExpt}. With all above improvements implemented, the performance of the multi-channel squeezed light array should reach 5 dB or higher squeezing in each channel.

This proof-of-principle demonstration of spatially-multiplexed squeezed light provides a scalable way to produce multiple squeezed light beams, which would be challenging for nonlinear systems requiring external cavities. Such squeezed light array can be applied in quantum imaging and sensing with high-spatial resolution~\cite{Bowen}, quantum network with more nodes~\cite{Kimble} and multi-partite entanglement studies~\cite{PKLam}. Although demonstrated here in the continuous variable regime, our scheme can be readily extended to discrete variable regime, i.e., to generate array of entangled photon pairs since entangled photon pair has already been demonstrated in such coated cells using hyperfine ground state coherence~\cite{Du}. Furthermore, it is possible to make the beams in the array correlate with each other at a quantum level, if a proper light configurations is used, as proposed in~\cite{HS}, thus allowing investigations of nonlinear photon-photon interaction~\cite{Chang} mediated by flying atoms and the future realization of entangled light arrays. This strategy can be extended to other systems such as trapped ions where the spin degree of freedom of the ions (encoded in the two lowest hyperfine states) are coupled via phonon-mediated long-range interactions induced by laser forces \cite{Blatt,Monroe,MSgate}.

The authors are grateful to Eugene S. Polzik for fruitful discussions, and Yanqiang Guo for drawing the experiment schematics.

\end{document}